\documentclass[10pt,twocolumn,letterpaper]{article}

\PassOptionsToPackage{table}{xcolor}

\usepackage[pagenumbers]{cvpr} 

\usepackage{xspace}
\newcommand{\model}{\textsc{LMEG-Doc}\xspace}
\usepackage{xcolor} 
\usepackage{booktabs}
\usepackage{graphicx}
\usepackage{subcaption}

\definecolor{cvprblue}{rgb}{0.21,0.49,0.74}
\usepackage[pagebackref,breaklinks,colorlinks,allcolors=cvprblue]{hyperref}

\def\confName{CVPR}
\def\confYear{2027}

\title{LMEG-Doc: Layout-Grounded Multi-View Evidence Graphs for Multimodal Long-Document Retrieval}

\author{
Changcheng Fu$^{1,2}$\thanks{Changcheng Fu and Xiao Liu contributed equally. \\
Work conducted during Changcheng Fu's internship at Capital One.} \quad
Xiao Liu$^{1,*}$ \quad
Qi Yu$^1$ \quad
Yage Wang$^1$ \quad
Derek Young$^1$ \quad
Giri Iyengar$^1$ \\[0.5em]
$^1$Capital One \\
$^2$University of Southern California
}

\begin{document}
\maketitle

\begin{abstract}
Retrieving relevant evidence from multimodal long
documents remains challenging, as supporting
information is often distributed across different
pages and modalities. Existing retrieval methods
primarily rank document elements by their individual
similarity to a question, often overlooking
relationships among complementary evidence.
Consequently, relevant figures and tables may
receive low retrieval scores even when they
contain essential information. To address this
limitation, we introduce \model, a layout-grounded
multi-view evidence graph framework for multimodal
long-document retrieval. We first construct an
evidence graph that captures structural and
contextual relationships among document elements,
and introduce Cross-Page Relation Refinement (CRR)
to discover additional connections across distant
pages through VLM-based verification. Building on
this graph, we combine multimodal retrieval signals
with graph-guided relevance propagation to recover
connected supporting evidence at both layout and
page levels. Experiments on MMDocIR, DocScope,
and FinRAGBench-V demonstrate consistent retrieval
improvements, including outperforming retrievers
fine-tuned on MMDocIR without task-specific training.
Furthermore, augmenting existing question-answering
pipelines with our retrieved evidence significantly improves answer accuracy with Qwen3.8-27B and outperforming the latest proprietary model including GPT-6.1 Sol, demonstrating the effectiveness
of relationship-aware multimodal retrieval.
\end{abstract}

\section{Introduction}
\label{sec:intro}

\begin{figure*}[t]
    \centering
    \includegraphics[width=0.95\textwidth]{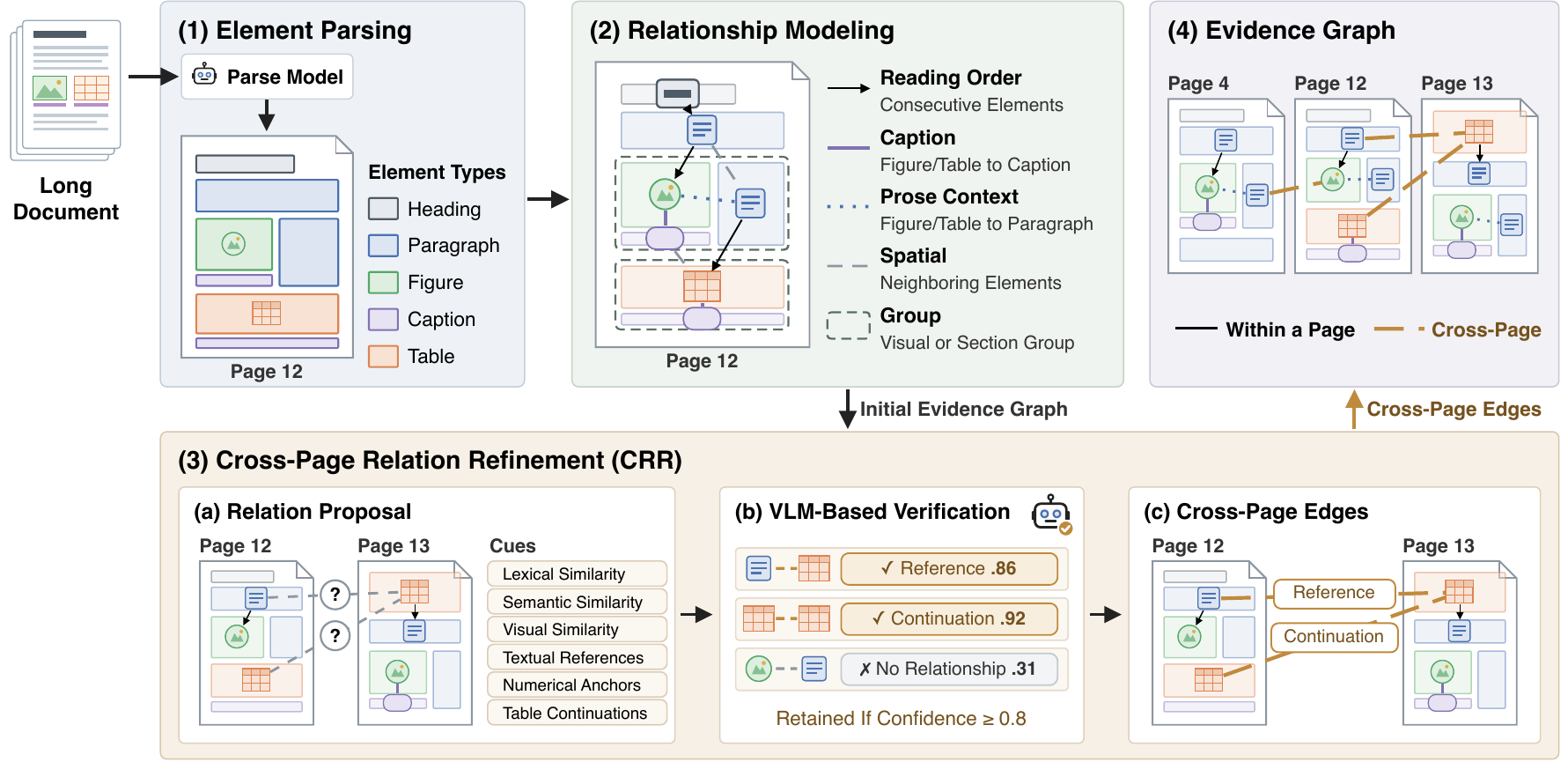}
    \caption{Evidence graph construction in \model.
    (1) Document element parsing converts each page into
    spatially grounded elements with types and bounding boxes.
    (2) Document relationship modeling connects elements within a
    page through reading-order, caption, prose-context, and spatial
    edges, and organizes them into visual and section groups.
    (3) Cross-Page Relation Refinement (CRR) proposes candidate
    pairs across pages from complementary cues and incorporates
    high-confidence, VLM-verified connections as cross-page edges.
    (4) The resulting evidence graph captures both local document
    structure and cross-page connections.}
    \label{fig:overview}
\end{figure*}

Multimodal long documents, such as scientific papers,
technical reports, and financial disclosures, contain
rich information distributed across textual passages,
figures, tables, and multiple pages. Retrieving relevant
evidence from these documents is essential for a wide
range of document understanding tasks, particularly
question answering that requires integrating information
across different modalities and
locations~\cite{mmlongbench,longdocurl,docscope}. However,
the diversity of document layouts and the dependencies
between distant elements make accurate multimodal
retrieval challenging.

\looseness=-1 To resolve the challenge, recent advances in multimodal document retrieval have
made substantial progress in identifying relevant
document content. Visual retrieval methods, such as
ColPali~\cite{colpali} and VisRAG~\cite{visrag},
directly represent document pages as images to preserve
visual and structural information. M3DocRAG~
\cite{m3docrag} extends multimodal retrieval to
complex multi-page and multi-document settings, while
MMDocIR~\cite{mmdocir} further highlights the
importance of retrieving evidence at both page and
layout-element levels. Agentic frameworks, such as
MDocAgent~\cite{mdocagent} and ViDoRAG~\cite{vidorag},
additionally explore iterative
retrieval for document question answering. Despite
these advances, existing retrieval approaches largely
rely on the individual similarity between a question
and a document page or element, without explicitly
modeling relationships among the retrieved evidence.

\looseness=-1 This reliance on individual similarity becomes
problematic when the document element most
relevant to a question does not directly contain
the answer. For example, a figure caption may
closely match the question, while the associated
figure contains the actual answer. Similarly,
a matching paragraph may refer to a table located
several pages away, where the required
information is presented. Although these
figures and tables contain essential evidence,
they may receive low retrieval scores due to
their limited direct similarity to the question
and consequently be overlooked. As a result,
independently ranking document elements can
produce fragmented evidence, failing to recover
the complementary information needed for
reliable question answering.

To overcome this limitation, retrieval should
consider not only the relevance of individual
document elements but also the relationships
among them. In fact, long documents naturally
contain structural connections, such as reading
order, captions, spatial proximity, and shared
sections, that associate related textual and
visual elements. Beyond these local connections,
semantic dependencies may further link figures,
tables, and passages across distant pages.
By explicitly modeling these relationships,
retrieval can move beyond independent similarity
matching to recover complementary evidence
that may receive low retrieval scores despite
being essential to answering the question.

\looseness=-1 Motivated by this insight, we introduce \model,
a layout-grounded multimodal retrieval framework
that organizes document elements into an evidence
graph and exploits their relationships for
question-conditioned retrieval. We first parse
documents into spatially grounded elements,
including text, figures, and tables, and construct
an evidence graph using naturally occurring
structural and contextual relationships.
To capture dependencies beyond individual pages,
we introduce Cross-Page Relation Refinement (CRR),
which proposes potential connections using
multimodal retrieval cues and verifies them
with a vision-language model. Together, these
relations provide a unified representation
of local document structure and long-range
evidence dependencies. The overview of \model is shown in Figure~\ref{fig:overview}.

\looseness=-1 Building on this graph, \model performs
retrieval at both layout and page levels.
At the layout level, we first combine lexical,
semantic, and visual retrieval signals to
identify directly relevant elements, then
propagate their relevance through the graph
to recover connected supporting evidence.
At the page level, we combine
whole-page visual similarity with
graph-derived element relevance to identify
pages containing complementary evidence.
The resulting page and layout rankings
enable flexible evidence selection and
can augment existing question-answering
pipelines without modifying their
underlying reasoning models.

To demonstrate the effectiveness of \model, made comprehensive evaluation on three multimodal
long-document benchmarks: MMDocIR\cite{mmdocir}, DocScope\cite{docscope},
and FinRAGBench-V\cite{finragbenchv}. Without task-specific
training, \model outperforms existing retrievers,
including MMDocIR-fine-tuned models, achieving
70.7\% layout recall at L@10.
These retrieval gains also improve downstream QA accuracy by 7.6\% on DocScope and 15.8\% on FinRAGBench-V with open-sourced Qwen3.8-27B model~\cite{qwen38}, outperforming the latest proprietary model and demonstrating the effectiveness of relationship-aware evidence retrieval.

Our main contributions are summarized as follows:
\begin{itemize}
    \item We introduce \model, a layout-grounded
    evidence graph that models structural and
    contextual relationships among multimodal
    document elements.

    \item \looseness=-1 Within \model, we propose Cross-Page Relation Refinement
    (CRR) and graph-guided multi-view retrieval
    to recover complementary evidence at both
    layout and page levels.

    \item We demonstrate consistent retrieval
    improvements across three long-document
    benchmarks and improved downstream
    document-grounded question answering.
\end{itemize}

\section{Related Work}
\label{sec:related}

\paragraph{Multimodal Document Retrieval.}\looseness=-1 
Text-based methods apply retrievers such as
BM25~\cite{bm25}, DPR~\cite{dpr},
ColBERT~\cite{colbert}, and BGE-M3~\cite{bge}
to OCR or parsed text, often losing visual
and layout information. Visual retrievers
instead operate directly on document
images~\cite{dse,colpali,visrag}, achieving
strong performance on benchmarks such as
ViDoRe~\cite{vidore3} and
MMDocIR~\cite{mmdocir}. Recent approaches
further explore region-level retrieval,
layout-aware indexing, and multimodal
fusion~\cite{regionrag,lfrag,midr,visdom}.
While these methods primarily rank pages
or regions by individual relevance,
\model additionally exploits relationships
among document elements to recover
complementary evidence.

\paragraph{Graph-Based Retrieval.}
Graph-based RAG organizes information
into structured representations for
multi-hop retrieval~\cite{graphragsurvey,graphrag,lightrag,raptor,hypergraphrag,gretriever}.
HippoRAG~\cite{hipporag,hipporag2}, for
example, employs personalized PageRank
to propagate relevance through knowledge
graphs. Related work explores document
layout graphs~\cite{layoutlmv3,docgraphlm,hegel,chdg,graph4mm},
entity-based multimodal
graphs~\cite{mmgraphrag,raganything,megarag,hvmgraphrag,mmgraphragvrd},
and graph-based retrieval over document
pages, chunks, or elements~\cite{molorag,eviprop,lilac,mldocrag,ladrag,magerag,docnavrag,hyperevidence,doctrace}.
In contrast, \model constructs a reusable
graph of spatially grounded document
elements, combining structural relationships
with VLM-verified cross-page connections.
It then propagates multi-view retrieval
relevance through the graph to produce
both page- and layout-level rankings
without task-specific training or
LLM inference during retrieval.

\paragraph{Long-Document Understanding.}
Document question answering has evolved
from single-page~\cite{docvqa} to
multi-page and long-document
settings~\cite{mpdocvqa,slidevqa,mmlongbench,longdocurl},
with recent benchmarks emphasizing
evidence grounding~\cite{docscope,finragbenchv}.
Existing methods either process complete
documents using VLMs~\cite{qwen25vl,qwen3vl,docopilot,docslm,urag,docseeker}
or employ retrieval-augmented
pipelines~\cite{rag,m3docrag,visrag},
including agentic frameworks that
iteratively retrieve and reason over
evidence~\cite{mdocagent,vidorag,simpledoc,doclens,mardoc,docdancer}.
Unlike approaches focused on improving
the answering process, \model enhances
evidence retrieval and can be integrated
into existing QA pipelines without
modifying their reasoning models.

\section{\model}
\label{sec:method}

\begin{figure*}[t]
    \centering
    \includegraphics[width=0.90\textwidth]{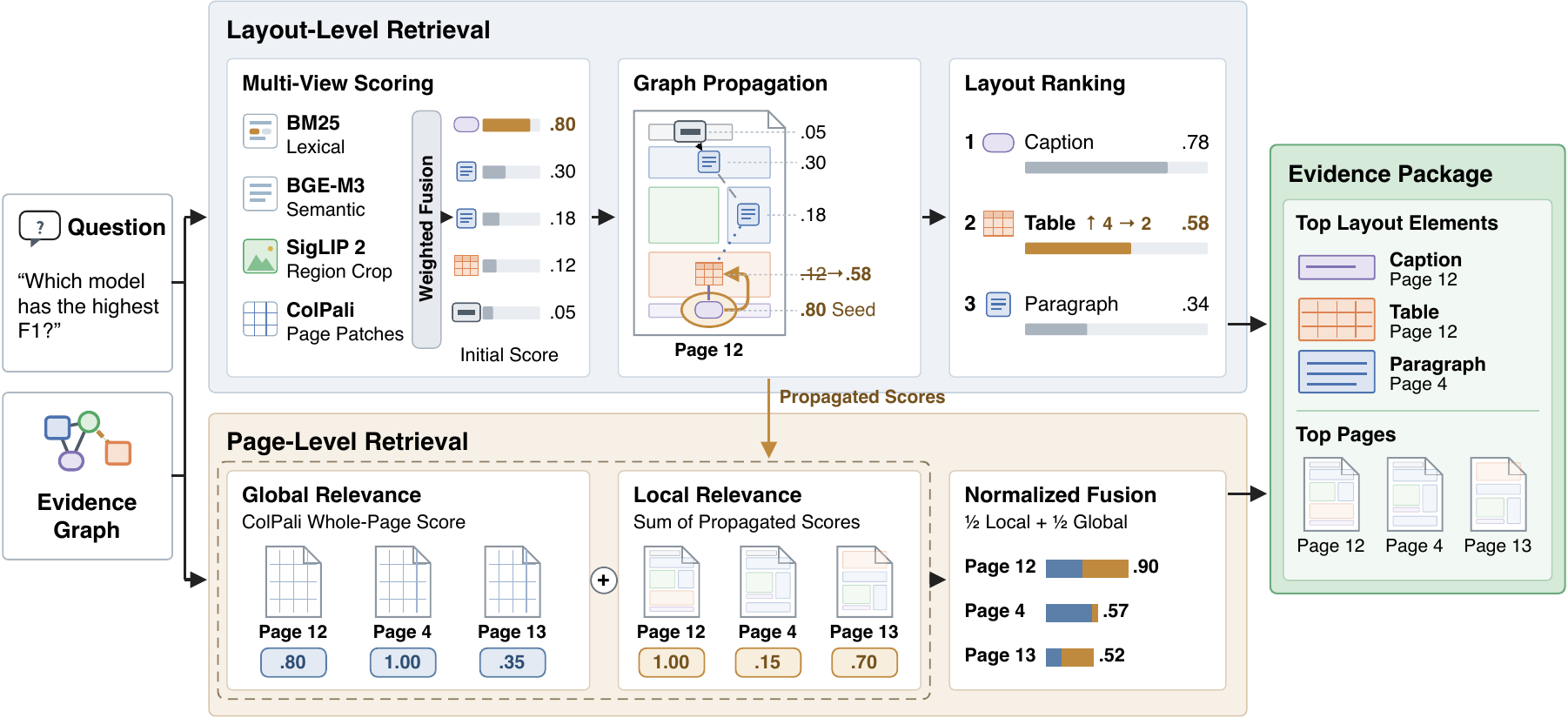}
    \caption{Graph-guided multimodal retrieval in \model.
    \emph{Layout-level retrieval} fuses four retrieval views into
    initial element scores, propagates relevance over the evidence
    graph with personalized PageRank, and ranks layout regions with
    reciprocal rank fusion and intra-group refinement: a caption
    that matches the question lifts its low-scoring table from
    rank 4 to rank 2. \emph{Page-level retrieval} fuses each page's
    normalized global relevance (ColPali whole-page score) and
    local relevance (sum of its propagated element scores). The
    top layout regions and pages form the evidence package.}
    \label{fig:retrieval}
\end{figure*}

\looseness=-1 In this section, we introduce \model, a
layout-grounded multimodal retrieval framework
that exploits relationships among document
elements to recover relevant and complementary
evidence from long documents. As illustrated
in Figure~\ref{fig:overview}, we first parse
a document $D$ into spatially grounded elements,
including text, figures, and tables, and
construct an evidence graph using naturally
occurring document relationships and
VLM-verified cross-page connections.

Given a question $q$, \model performs graph-guided retrieval at both layout and page levels, as shown in Figure~\ref{fig:retrieval}. At the layout level, we fuse four multimodal signals to retrieve relevant elements and propagate relevance through the graph to identify supporting evidence. At the page level, we combine whole-page visual relevance with graph-derived element relevance to retrieve pages containing complementary evidence. The resulting rankings form a flexible evidence package that integrates with existing question-answering pipelines without modifying their reasoning models. We detail each component in the following subsections.

\subsection{Evidence Graph Construction}
\label{sec:method:construction}

\looseness=-1 To capture relationships among document elements,
we first construct a question-independent evidence
graph that can be reused across queries. This process
begins with document parsing, followed by establishing
structural relationships and discovering implicit
cross-page connections, as demonstrated in Figure~\ref{fig:overview}.

\paragraph{Document Element Parsing.}
We first parse each document into individual elements with Nemotron-Parse v1.2~\cite{nvidia2026nemotronparse},
including paragraphs, headings, captions, figures,
tables, and other layout components. To preserve
their original document context, each element is
represented as
\begin{equation}
    v_i = (p_i, b_i, t_i, x_i),
\end{equation}
where $p_i$ denotes the page index, $b_i$ is the
normalized bounding box, $t_i$ specifies the element
type, and $x_i$ contains the extracted text when
available. By preserving
both element types and spatial locations, this
representation provides the foundation for
constructing document relationships and enables
precise evidence retrieval and citation.


\paragraph{Document Relationship Modeling.}\looseness=-1 
Building on the parsed document elements, we construct
an evidence graph $\mathcal{G}=(\mathcal{V},\mathcal{E})$,
where each node represents a document element and
each edge captures a relationship between elements.
We first establish four types of direct relationships:
1) reading-order edges connect consecutive elements
within a page; 2) caption edges associate figures
and tables with their captions; 3) prose-context
edges link visual elements to surrounding paragraphs;
and 4) spatial edges connect neighboring elements
based on their bounding-box proximity.

Beyond pairwise relationships, we introduce groups
to represent collections of document elements that
share a common semantic or structural context.
Specifically, a visual group contains a figure,
table, or equation together with its associated
caption and explanation, while a section group contains a heading
and the elements belonging to that section.
Each group is represented as an auxiliary node,
connected to its constituent elements through
membership edges. This allows retrieval
to capture related elements through their shared
context, even when only one member directly matches
the question. Each edge records its relation type
and confidence, providing an interpretable structure
for subsequent graph-guided retrieval.

\paragraph{Cross-Page Relationship Refinement.}\looseness=-1 
The relationships established above primarily capture
intra-page document structure, but supporting evidence may
also be connected across distant pages.
To capture these dependencies, we introduce Cross-Page
Relation Refinement (CRR), which extends the initial
evidence graph through relation proposal and VLM-based
verification.

In the proposal stage, we identify candidate pairs
across different pages using complementary cues,
including lexical and semantic similarity, visual
similarity, textual references, shared numerical
anchors, and potential table continuations. We
restrict candidates to pairs containing at least
one visual element to maintain multimodal connectivity. The details 
of each cues are listed in supplementary meterial Section~\ref{sec:supp:crr_proposal}.

In the verification stage, a VLM examines each
candidate pair using the corresponding image regions,
textual content, page indices, and element types.
It determines whether a meaningful relationship
exists and predicts its type and confidence.
High-confidence connections are incorporated into
the evidence graph as cross-page edges, enabling
retrieval to discover related information beyond
local document structure. Additional implementation
details are provided in the supplementary material Section~\ref{sec:supp:crr_verification}.

\noindent Together, the layout-grounded relationships and
verified cross-page connections form a unified
evidence graph that captures both local document
structure and long-range multimodal dependencies.
By preserving these relationships, \model enables
subsequent retrieval to move beyond directly
matching elements and discover complementary
supporting evidence within and across pages.

\subsection{Graph-Guided Multimodal Retrieval}
\label{sec:method:retrieval}

Building on the constructed evidence graph, we perform
question-conditioned retrieval demonstrated in Figure~\ref{fig:retrieval}
to identify both directly
relevant document elements and their connected supporting
evidence.

\paragraph{Layout-Level Retrieval.}
We first estimate the relevance of individual parsed
elements using four complementary retrieval views.
BM25~\cite{bm25} captures lexical matches between
the question and element text, while BGE-M3~\cite{bge}
measures their semantic similarity through text
embeddings. For visual evidence, SigLIP 2~\cite{siglip}
computes the similarity between the question and
cropped document regions. ColPali~\cite{colpali}
further captures visual relevance within the
context of the original page through patch-level
late interaction.

Specifically, for each evidence unit $v_i$, we
compute its region-level ColPali score as
\begin{equation}
s_{\mathrm{ColPali}}(q,v_i)
=
\sum_{a=1}^{L_q}
\max_{j\in\mathcal{S}(b_i)}
Q_a^\top Z_{p_i,j},
\label{eq:colpali}
\end{equation}
where $L_q$ denotes the number of query tokens,
$Q_a$ is the embedding of the $a$-th query token,
and $Z_{p_i,j}$ represents the $j$-th visual patch
embedding of page $p_i$. The set $\mathcal{S}(b_i)$
contains the patches within the bounding box of
$v_i$.

Since these views operate on different scales,
we normalize their scores independently and combine them through
weighted fusion:
\begin{equation}
z_m(i)=
\frac{s_m(q,v_i)-\mu_m}{\sigma_m},
\quad
s(i)=\sum_{m\in\mathcal{M}}w_m z_m(i),
\label{eq:fusion}
\end{equation}
\looseness=-1 where $\mu_m$ and $\sigma_m$ denote the median
and scaled median absolute deviation of the
scores produced by view $m$, respectively.
$w_m$ controls the contribution of each view.
The resulting $s(i)$ provides the initial
relevance score of evidence unit $v_i$.

\looseness=-1 While multi-view scoring identifies elements that
directly match the question, it may overlook related
evidence with low retrieval scores. To recover such
information, we propagate relevance through the
evidence graph using personalized PageRank~\cite{ppr}.
Specifically, we select the top-$K$ elements as
initial seeds and construct a bounded subgraph
around them. Relevance is then propagated along
graph edges according to their relation types
and confidence scores, allowing connected elements
to receive relevance from the initial seeds.
To prevent frequent spatial connections from
dominating propagation, we balance transitions
across different relation types.

The propagation process is defined as
\begin{equation}
r^{(\ell+1)}
=
(1-\alpha)r_0
+
\alpha T^\top r^{(\ell)},
\label{eq:ppr}
\end{equation}
where $r_0$ is the initial relevance distribution
over seed elements, $T$ is the relation-aware
transition matrix, and $\alpha$ controls the
propagation strength. This enables retrieval
to discover supporting figures, tables, and
passages that may not directly match the question
but are connected to relevant evidence. The construction of transition matrix is listed in supplementary material Section~\ref{sec:supp:propagation}.

The resulting graph relevance scores allow retrieval
to identify related figures, tables, and passages
that may have low direct retrieval scores but contain
complementary supporting information. To preserve
strong direct matches while incorporating graph-discovered
evidence, we combine the initial retrieval ranking
and the graph-propagated ranking using reciprocal
rank fusion (RRF)~\cite{rrf}:
\begin{equation}
s_{\mathrm{RRF}}(i)
=
\frac{1}{\kappa+\operatorname{rank}_{s}(i)}
+
\frac{1}{\kappa+\operatorname{rank}_{r}(i)},
\label{eq:rrf}
\end{equation}
where $\operatorname{rank}_{s}(i)$ and
$\operatorname{rank}_{r}(i)$ denote the ranks of
element $v_i$ under the initial retrieval scores
and graph-propagated relevance, respectively, and
$\kappa$ is a smoothing constant. This fusion
preserves directly relevant elements while promoting
supporting evidence discovered through graph traversal.

Finally, we refine relevance scores within the
visual and section groups defined during graph
construction. For visual groups, all members
are promoted to the highest score within their
group, ensuring that associated figures, tables,
and captions are ranked together. For section
groups, we selectively promote the heading,
the highest-scoring element and its neighbors,
and other highly relevant members. This
refinement preserves contextual relationships
without introducing excessive irrelevant
content, producing the final layout-level ranking. We detail the formal intra-group score refinement process in supplementary material Section~\ref{sec:supp:group_refinement}.

\paragraph{Page-Level Retrieval.}\looseness=-1 
Building on the layout-level retrieval results,
we compute page relevance using two complementary
signals: global visual relevance and local evidence
relevance. The global signal captures how well
an entire page matches the question, while the
local signal reflects the relevance of its
constituent elements and their connections.

For global relevance, we use ColPali~\cite{colpali}
to compute the whole-page relevance score
$S_{\mathrm{global}}(p)$ through standard
patch-level late interaction. For local relevance, we aggregate the
graph-propagated scores of elements belonging
to each page:
\begin{equation}
S_{\mathrm{local}}(p)
=
\sum_{v_i\in\mathcal{V}_p} r_i,
\label{eq:page_aggregation}
\end{equation}
\looseness=-1 where $\mathcal{V}_p$ denotes the set of document
elements on page $p$, and $r_i$ is the
graph-propagated relevance score of element $v_i$.

We combine the global and local relevance scores
through normalized score fusion:
\begin{equation}
S_{\mathrm{page}}(p)
=
\eta\,\widehat{S}_{\mathrm{local}}(p)
+
(1-\eta)\,\widehat{S}_{\mathrm{global}}(p),
\label{eq:page}
\end{equation}
where $\widehat{S}_{\mathrm{local}}(p)$ and
$\widehat{S}_{\mathrm{global}}(p)$ denote the
min--max normalized local and global relevance
scores across document pages, respectively.
We set $\eta=0.5$ to balance their contributions.
The resulting score determines the final page
ranking without modifying the layout-level ranking.


\paragraph{Evidence Package.}\looseness=-1 
With both page-level and layout-level relevance scores,
\model supports flexible evidence selection according
to downstream task requirements to form the evidence package. For page-level retrieval, we select the highest-ranked pages. For element-level tasks, evidence is selected based on layout relevance, optionally combined with page relevance to preserve broader context. Each retrieved element retains its page index, bounding box, and type for precise localization. This allows \model to provide evidence at different granularities for multimodal document understanding.

\begin{table}[t]
    \centering
    \caption{Benchmark datasets and evaluation tasks.
    P@$k$ denotes page recall, and L@$k$ denotes
    layout-level evidence recall at top-$k$.}
    \label{tab:benchmarks}
    \small
    \setlength{\tabcolsep}{5pt}
    \begin{tabular}{l|ccc}
        \toprule
        \textbf{Benchmark} & \textbf{Questions}
        & \textbf{Retrieval} & \textbf{QA} \\
        \midrule
        MMDocIR~\cite{mmdocir}
        & 1,658 & P@$k$, L@$k$ & -- \\
        DocScope~\cite{docscope}
        & 730 & P@$k$, L@$k$ & Accuracy \\
        FinRAGBench-V~\cite{finragbenchv}
        & 539 & P@$k$, L@$k$ & Accuracy \\
        \bottomrule
    \end{tabular}
    \vspace{-2mm}
\end{table}

\subsection{Integration with Question Answering}
\label{sec:method:qa}

\looseness=-1 Once the evidence package is obtained, we integrate
it into existing question-answering pipelines as
additional context, without modifying the underlying
reasoning model or answering strategy. This enables
\model to support both direct VLM answering and
agentic question-answering frameworks.
This plug-and-play design allows \model to enhance
existing multimodal question-answering pipelines
through relationship-aware evidence retrieval,
without requiring changes to their reasoning
or answer-generation mechanisms.

\section{Experiments}
\label{sec:experiments}

We evaluate \model on three multimodal long-document
benchmarks, covering both evidence retrieval and
document-grounded question answering (QA), as
summarized in Table~\ref{tab:benchmarks}.
We first assess its ability to identify relevant
pages and layout elements, then examine whether
the retrieved evidence improves downstream
answering performance across different document
domains and QA settings.

\subsection{Benchmark Datasets and Evaluation Metrics}
\label{sec:datasets}

\paragraph{Evidence Retrieval.}\looseness=-1 
MMDocIR~\cite{mmdocir} contains 1,658 expert-annotated questions
over 313 documents spanning ten domains, with
ground-truth pages and layout regions.
Following its evaluation protocol, we report
page recall (P@$k$), measuring the fraction
of ground-truth pages retrieved among the top-$k$
results, and layout-level recall (L@$k$),
measuring the fraction of ground-truth region
area covered by the top-$k$ retrieved elements
on the gold pages. To assess the
robustness and generalizability of \model across
benchmarks, we further apply the same metrics
to DocScope~\cite{docscope} and
FinRAGBench-V~\cite{finragbenchv} using their
available evidence annotations.

\paragraph{Document QA.}\looseness=-1 
We evaluate downstream question answering on
DocScope and FinRAGBench-V following their
respective evaluation protocols. DocScope
provides 730 test questions in total.For FinRAGBench-V, we
use its English subset of 539 questions over
financial reports represented as page images.
Following the official protocols, we employ
LLM-as-a-judge to assess answer correctness.
Additional evaluation details are provided
in the supplementary material in Section~\ref{sec:supp:evaluation}.

\subsection{Implementation Details}
\label{sec:implementation}

We use NVIDIA Nemotron-Parse v1.2~\cite{nvidia2026nemotronparse}
for document parsing and four retrieval models:
BM25~\cite{bm25}, BGE-M3~\cite{bge},
SigLIP 2~\cite{siglip}, and ColPali v1.3~\cite{colpali}.
For cross-page relation verification, we use
Qwen3.6-35B-A3B~\cite{qwen36} on all the datasets. We use Qwen3.8-27B
for downstream question answering and
Qwen3.6-35B-A3B for answer evaluation.
Further details on graph construction,
cross-page link verification, retrieval
configuration, rendering, and evaluation
are provided in the supplementary material
Sections~\ref{sec:supp:graph_construction}--\ref{sec:supp:evaluation}.

\begin{table}[t]
\centering
\caption{Retrieval results on MMDocIR.
P@$k$ and L@$k$ denote page- and layout-level
recall, respectively. $^\dagger$ indicates
training on MMDocIR.}
\label{tab:sota_mmdocir}
\small

\resizebox{0.9\linewidth}{!}{%
\begin{tabular}{l|ccc|ccc}
\toprule
& \multicolumn{3}{c}{\textbf{Page}}
& \multicolumn{3}{|c}{\textbf{Layout}} \\
\cmidrule(lr){2-4}
\cmidrule(lr){5-7}
\textbf{Method} & P@1 & P@3 & P@5
       & L@1 & L@5 & L@10 \\
\midrule

\rowcolor{black!8}
\multicolumn{7}{l}{\bfseries Text-based Retrieval} \\
ColBERT (OCR)~\cite{colbert}
& 40.3 & 58.8 & 66.0
& 19.8 & 33.2 & 37.6 \\
ColBERT (VLM text)~\cite{colbert}
& 45.8 & 64.9 & 72.3
& 31.3 & 54.4 & 61.9 \\

\addlinespace[2pt]
\rowcolor{black!8}
\multicolumn{7}{l}{\bfseries Visual-based Retrieval} \\
DSE (Docmatix)~\cite{dse}
& 50.2 & 71.4 & 79.5
& 27.9 & 49.8 & 57.5 \\
ColPali~\cite{colpali}
& 53.0 & 74.7 & 80.8
& \textbf{32.7} & 54.0 & 62.0 \\
DPR-Phi3$^\dagger$~\cite{mmdocir}
& 54.1 & 73.5 & 81.1
& 29.5 & 51.6 & 60.2 \\
Col-Phi3$^\dagger$~\cite{mmdocir}
& \textbf{57.0} & 76.3 & 82.2
& 31.1 & 52.3 & 61.1 \\

\addlinespace[2pt]
\rowcolor{black!8}
\multicolumn{7}{l}{\bfseries Ours} \\
\textbf{\model}
& 56.7 & \textbf{80.3} & \textbf{85.7}
& 32.1 & \textbf{61.1} & \textbf{70.7} \\

\bottomrule
\end{tabular}%
}
\end{table}

\subsection{Evidence Retrieval Results}
\label{sec:exp:retrieval}

\noindent \looseness=-1 \textbf{\model outperforms existing
retrievers, including those fine-tuned on MMDocIR.}
We first compare \model with published text-based
and visual retrievers on MMDocIR, including
ColBERT~\cite{colbert}, DSE~\cite{dse},
ColPali~\cite{colpali}, and the MMDocIR-trained
DPR-Phi3 and Col-Phi3 models~\cite{mmdocir}.

\begin{table*}[!ht]
\centering
\caption{Retrieval performance on DocScope and
FinRAGBench-V. P@$k$ denotes page recall and
L@$k$ denotes layout-level recall.
Best results are highlighted in bold. ColPali method ranks only for whole page therefore layout performance is not applicable.}
\label{tab:retrieval}
\small
\resizebox{0.8\textwidth}{!}{%
\begin{tabular}{l|ccc|ccc|ccc|ccc}
\toprule
& \multicolumn{6}{c|}{\textbf{DocScope}}
& \multicolumn{6}{c}{\textbf{FinRAGBench-V}} \\
\cmidrule(lr){2-7}
\cmidrule(lr){8-13}
& \multicolumn{3}{c|}{\textbf{Page}}
& \multicolumn{3}{c|}{\textbf{Layout}}
& \multicolumn{3}{c|}{\textbf{Page}}
& \multicolumn{3}{c}{\textbf{Layout}} \\
\cmidrule(lr){2-4}
\cmidrule(lr){5-7}
\cmidrule(lr){8-10}
\cmidrule(lr){11-13}
\textbf{Method}
& P@1 & P@3 & P@5
& L@1 & L@5 & L@10
& P@1 & P@3 & P@5
& L@1 & L@5 & L@10 \\
\midrule
BM25~\cite{bm25}
& 28.4 & 51.6 & 60.5
& 9.8 & 21.1 & 26.9
& 43.9 & 60.1 & 66.4
& 8.4 & 16.6 & 18.3 \\
BGE-M3~\cite{bge}
& 29.6 & 51.9 & 62.5
& 12.1 & 28.4 & 36.9
& 35.3 & 52.6 & 57.3
& 5.9 & 12.4 & 16.9 \\
ColPali~\cite{colpali}
& \textbf{43.4} & 67.1 & 75.9
& -- & -- & --
& \textbf{61.8} & 78.1 & 82.5
& -- & -- & -- \\
\midrule
\textbf{\model}
& 40.2 & \textbf{67.1} & \textbf{76.0}
& \textbf{16.8} & \textbf{44.1} & \textbf{57.1}
& 57.9 & \textbf{80.5} & \textbf{86.0}
& \textbf{15.3} & \textbf{44.1} & \textbf{52.0} \\
\bottomrule
\end{tabular}%
}
\end{table*}

As shown in Table~\ref{tab:sota_mmdocir},
\model achieves the best page-level recall
at P@3 and P@5 and layout-level recall
at L@5 and L@10, without any task-specific
training. Notably, \model surpasses
Col-Phi3, a retriever trained on MMDocIR,
by 4.0\% and 3.5\%
at P@3 and P@5, respectively, while achieving
comparable performance at P@1.
It also substantially improves layout-level
retrieval over the strongest published
baseline, increasing L@10 from 62.0\%
to 70.7\%. These results suggest that
explicitly modeling relationships between
document elements can recover complementary
evidence more effectively than relying
on learned retrieval similarity alone.

\noindent \textbf{Relationship-aware retrieval
generalizes across benchmarks and is particularly
effective at larger retrieval budgets.}
We further evaluate \model on DocScope and
FinRAGBench-V against text-based BM25~\cite{bm25}, BGE-M3~\cite{bge}, and visual-based ColPali~\cite{colpali} retrieval methods.
As shown in Table~\ref{tab:retrieval},
\model achieves the best P@3, P@5, L@5,
and L@10 on both benchmarks, while maintaining
comparable P@1 performance.
Notably, on FinRAGBench-V, \model improves
P@5 by 3.5\% over ColPali
and achieves an L@10 of 52.0\%, substantially
outperforming the baseline retrievers.
These results suggest that graph-guided retrieval
effectively recovers complementary evidence
across diverse document domains, particularly
when multiple supporting elements are required.

\subsection{Document-Grounded QA Results}
\label{sec:exp:qa}


\begin{table}[t]
\centering
\caption{Overall answer accuracy (\%) on DocScope.}
\label{tab:sota_docscope}
\small
\resizebox{0.70\linewidth}{!}{
\begin{tabular}{l|c}
\toprule
\textbf{Method} & \textbf{Accuracy} \\
\midrule
\rowcolor{black!8}
\multicolumn{2}{l}{\bfseries Proprietary Models} \\
Gemini 3.1 Pro~\cite{gemini31pro} & 78.9 \\
Claude Opus 4.7~\cite{claudeopus47} & 76.0 \\
Claude Sonnet 4.6~\cite{claudesonnet46} & 70.5 \\
GPT-5.4~\cite{gpt54} & 62.7 \\
GPT-6.1 Sol~\cite{openai2026gpt61sol} & 78.4 \\
\addlinespace[2pt]
\rowcolor{black!8}
\multicolumn{2}{l}{\bfseries Open-Weight Models} \\
Qwen3.5-397B-A17B~\cite{qwen35} & 63.8 \\
Qwen3.5-27B~\cite{qwen35} & 66.4 \\
Qwen3-VL-235B-A22B~\cite{qwen3vl} & 51.4 \\
Gemma-4-31B~\cite{gemma4} & 59.4 \\
Qwen3.8-27B~\cite{qwen38} & 71.6 \\
\addlinespace[2pt]
\rowcolor{black!8}
\multicolumn{2}{l}{\bfseries Document QA Frameworks} \\
SimpleDoc~\cite{simpledoc} & 50.5 \\
ViDoRAG~\cite{vidorag} & 21.2 \\
URaG~\cite{urag} & 16.3 \\
Docopilot-8B~\cite{docopilot} & 7.4 \\
\addlinespace[2pt]
\rowcolor{black!8}
\multicolumn{2}{l}{\bfseries Ours} \\
\textbf{Qwen3.8-27B + \model}
& \textbf{79.2} \\
\bottomrule
\end{tabular}

}

\end{table}

\noindent \looseness=-1 \textbf{Open-sourced model adding \model evidence package achieves competitive
performance with leading proprietary models.}
As shown in Table~\ref{tab:sota_docscope},
Qwen3.8 + \model achieves 79.2\% overall accuracy on
DocScope, surpassing the proprietary model Gemini 3.1 Pro~\cite{gemini31pro}.
Notably, this setting uses an light open-weight Qwen3.8-27B\cite{qwen38}
answering model without task-specific fine-tuning,
demonstrating the effectiveness of retrieval-based
evidence augmentation.

\begin{figure}[!t]
    \centering
    \includegraphics[width=0.8\linewidth]{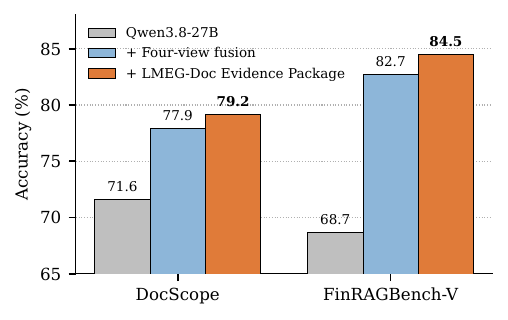}
    \caption{Impact of evidence augmentation on document QA.}
    \label{fig:qa}
\end{figure}

\noindent \looseness=-1 \textbf{Relationship-aware evidence
retrieval substantially improves downstream QA.}
As illustrated in Figure~\ref{fig:qa},
augmenting Qwen3.8-27B with \model improves
answer accuracy by 7.6\%
on DocScope and 15.8\% on FinRAGBench-V
over the original answering pipeline.
Compared with four-view fusion introduced in Equation~\ref{eq:fusion}, our augmented setting
further improves accuracy by 1.3\% and
1.8\%, respectively.
These results indicate that modeling
relationships among document elements helps
recover complementary supporting evidence
beyond multimodal similarity-based retrieval,
leading to more accurate answers across
different document domains.

\section{Analysis}
\label{sec:analysis}

\subsection{Ablation Studies}
\label{sec:exp:ablation}

\begin{table}[t]
\centering
\caption{Ablation study of \model on MMDocIR
using layout-level recall L@10 for all
questions and the multi-page subset.}
\label{tab:ablations}
\small
\resizebox{0.93\linewidth}{!}{%
\begin{tabular}{l|cc}
\toprule
\textbf{Method}
& \textbf{All}
& \textbf{Multi-page} \\
\midrule
\textbf{\model}
& \textbf{70.7} & \textbf{54.8} \\
$-$ Cross-Page Relation Refinement (CRR)
& 70.5 & 52.9 \\
$-$ Graph Propagation (PPR)
& 68.5 & -- \\
$-$ Intra-Group Score Refinement
& 69.6 & -- \\
\bottomrule
\end{tabular}%
}
\end{table}

\looseness=-1 To further assess the effectiveness of \model,
we evaluate the contribution of each component
on MMDocIR using layout-level recall L@10.
As shown in Table~\ref{tab:ablations},
removing graph propagation or intra-group
score refinement reduces L@10 by 2.2\%
and 1.1\%, respectively,
demonstrating their complementary contributions
to recovering supporting evidence.

\looseness=-1 Removing cross-page relation refinement (CRR)
results in a smaller decrease of 0.2\%
when evaluated over all questions.
However, when considering only multi-page
questions, its contribution becomes more
pronounced: CRR improves L@10 from 52.9\%
to 54.8\%.
This suggests that while page-local relationships
account for most of the overall retrieval gains,
verified cross-page connections are particularly
valuable for recovering evidence distributed
across multiple pages.
Further ablations on relation types and the
ColPali fusion weight can be found in
Sections~\ref{sec:supp:ablation} and~\ref{sec:supp:colpali_weight} of the supplementary material.

\subsection{Analysis by Evidence Modality}
\label{sec:exp:modality}
\begin{table}[t]
\centering
\caption{Retrieval and localization results by evidence modality.
Cross represents questions require cross-modal evidences.}
\label{tab:modality}
\footnotesize
\setlength{\tabcolsep}{2.3pt}
\resizebox{0.90\linewidth}{!}{
\begin{tabular}{@{}l| cc cc cc cc@{}}
\toprule
& \multicolumn{2}{c}{\textbf{Text}} & \multicolumn{2}{c}{\textbf{Table}} & \multicolumn{2}{c}{\textbf{Figure}} & \multicolumn{2}{c}{\textbf{Cross}} \\
\cmidrule(lr){2-3} \cmidrule(lr){4-5} \cmidrule(lr){6-7} \cmidrule(l){8-9}
\textbf{Method} & P@10 & L@10  & P@10  & L@10  & P@10  & L@10  & P@10  & L@10  \\
\midrule
BM25~\cite{bm25} & 84.1 & 61.1 & 73.9 & 43.0 & 71.7 & 25.6 & 74.1 & 27.2 \\
BGE-M3~\cite{bge} & 87.7 & 63.1 & 78.0 & 56.6 & 76.7 & 25.5 & 78.0 & 36.1 \\
ColPali~\cite{colpali} & 88.2 & -- & 84.8 & -- & 93.3 & -- & \textbf{90.6} & -- \\
\midrule
Four-view & 88.4 & 68.5 & 88.3 & 77.9 & 93.5 & 70.2 & 84.3 & 61.3 \\
\textbf{LMEG-Doc} & \textbf{91.6} & \textbf{70.6} & \textbf{88.4} & \textbf{78.8} & \textbf{94.4} & \textbf{77.6} & \textbf{90.6} & \textbf{64.3} \\
\bottomrule
\end{tabular}

}

\end{table}

\looseness=-1 We analyze retrieval performance across questions with text,
table, figure, and cross-modal ground-truth on
MMDocIR, as shown in Table~\ref{tab:modality}.
\model achieves the best layout-level recall
across all four modalities, with the largest
gain over four-view fusion on figure-based
questions by 7.4\% at L@10. For cross-modal
questions, \model further improves P@10
by 6.3\% and L@10 by 3.0\%.
These results highlight the effectiveness
of modeling relationships among document
elements, particularly for retrieving visual
and cross-modal evidence.
Results on DocScope and FinRAGBench-V can be
found in Section~\ref{sec:supp:modality}
of the supplementary material.

\subsection{Analysis by Evidence Distribution}
\label{sec:exp:multipage}
\begin{table}[t]
\centering
\caption{Retrieval performance on single-page
and multi-page questions in MMDocIR.
P@10 and L@10 denote page- and layout-level
recall, respectively.}
\label{tab:pages}
\small

\resizebox{0.8\linewidth}{!}{%
\begin{tabular}{l|cc|cc}
\toprule
& \multicolumn{2}{c|}{\textbf{Single-page}}
& \multicolumn{2}{c}{\textbf{Multi-page}} \\
\cmidrule(lr){2-3}
\cmidrule(lr){4-5}
\textbf{Method}
& P@10 & L@10 & P@10 & L@10 \\
\midrule
BM25~\cite{bm25}
& 80.8 & 50.7 & 65.3 & 19.4 \\
BGE-M3~\cite{bge}
& 85.1 & 55.8 & 68.5 & 25.1 \\
ColPali~\cite{colpali}
& 90.7 & -- & 81.1 & -- \\
\midrule
Four-view fusion
& 91.2 & 74.5 & 79.6 & 51.5 \\
\textbf{\model}
& \textbf{93.5} & \textbf{77.7}
& \textbf{81.9} & \textbf{54.8} \\
\bottomrule
\end{tabular}%
}
\end{table}

\begin{figure}[t]
    \centering
    \includegraphics[width=0.80\linewidth]{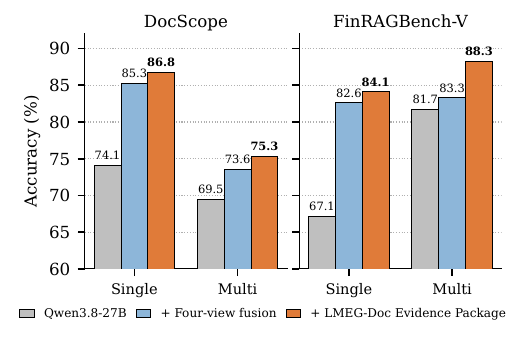}
    \caption{QA performance by evidence distribution.
    \model improves answer accuracy on both single-page
    and multi-page questions.}
    \label{fig:pages_accuracy}
\end{figure}
\looseness=-1 We further analyze the effectiveness of \model
on single-page and multi-page questions.
As shown in Table~\ref{tab:pages}, \model
achieves the highest page- and layout-level
recall on both subsets of MMDocIR.
Compared with four-view fusion, it improves
L@10 by 3.2\% and 3.3\%
on single-page and multi-page questions,
respectively, demonstrating the benefit
of modeling document relationships across
different evidence distributions.

These retrieval gains also translate into
improved downstream QA performance, as
illustrated in Figure~\ref{fig:pages_accuracy}.
Using the same answering model, \model
consistently outperforms four-view fusion
on both question types. Notably, on multi-page
questions, it improves answer accuracy
by 1.7\% and 5.0\%,
respectively. These results highlight
the effectiveness of relationship-aware
retrieval in supporting reasoning over
evidence distributed across multiple pages.
Retrieval results on DocScope and FinRAGBench-V
can be found in Section~\ref{sec:supp:multipage}
of the supplementary material.

\section{Conclusion}
\label{sec:conclusion}

In this work, we introduced \model, a
layout-grounded multimodal retrieval framework
that explicitly models relationships among
document elements to recover complementary
evidence in long documents. By combining
document structure, VLM-verified cross-page
relations, and graph-guided relevance propagation,
\model moves beyond independent similarity-based
retrieval to identify connected evidence across
pages and modalities. Experiments on MMDocIR,
DocScope, and FinRAGBench-V demonstrate consistent
improvements in page- and layout-level retrieval,
particularly for visual and cross-modal evidence.
Moreover, integrating the retrieved evidence
into existing question-answering pipelines
improves downstream answer accuracy without
modifying the underlying reasoning model.
These findings highlight the importance of
relationship-aware evidence retrieval for
reliable multimodal long-document understanding.

{
    \small
    \bibliographystyle{ieeenat_fullname}
    \bibliography{main}
}

\clearpage
\setcounter{page}{1}
\maketitlesupplementary
\setcounter{section}{0}
\renewcommand{\theHsection}{supp.\arabic{section}}

\ifdefined\lmegtblbox\else\newsavebox{\lmegtblbox}\fi
\providecommand{\fitbox}[2]{\sbox{\lmegtblbox}{#2}\ifdim\wd\lmegtblbox>#1 \resizebox{#1}{!}{\usebox{\lmegtblbox}}\else\usebox{\lmegtblbox}\fi}

\section{Evidence Graph Construction Details}
\label{sec:supp:graph_construction}

\subsection{Document Parsing and Alignment}
\label{sec:supp:parsing}

Each page is parsed by Nemotron-Parse
v1.2~\cite{nvidia2026nemotronparse} into elements
$v_i=(p_i,b_i,t_i,x_i)$ (Section~\ref{sec:method:construction}).
Element types include paragraphs, headings, captions,
tables, figures, equations, footnotes, and lists.
Tables keep the parser's markdown rendering as their
text $x_i$, whereas figures carry no text and are
represented by crops of the page image.

On MMDocIR, which provides an official inventory of
candidate layout regions, parsed elements are aligned
with the candidates using an intersection-over-union
threshold of $0.5$ or a containment threshold of $0.8$.
Evaluation therefore uses the benchmark's regions,
while the graph retains the parser's element types
and reading order. Unmatched captions, headings,
titles, and footnotes are retained as \emph{bridge
nodes}: they transmit relevance during propagation
but are excluded from the final ranking. On DocScope
and FinRAGBench-V, which do not provide a candidate
inventory, the parsed elements themselves serve as
retrieval candidates and are scored directly against
the ground-truth boxes.

\subsection{Document Relationships and Groups}
\label{sec:supp:relations}

All direct relationships in
Section~\ref{sec:method:construction} are produced by
fixed rules. Each edge records the rule that created it
and a confidence $c\in[0,1]$, so every retrieved
connection can be traced back to its origin.
\begin{itemize}
    \item \textbf{Reading-order} edges connect
    consecutive elements on a page.
    \item \textbf{Caption} edges attach a caption to
    the nearest visual element on its page.
    \item \textbf{Prose-context} edges connect a visual
    element with the paragraphs immediately before
    and after it.
    \item \textbf{Spatial} edges connect each element
    to its six nearest neighbors on the same page.
\end{itemize}
Caption and prose-context relations are established
using fixed distance thresholds. A visual group
contains a figure, table, or equation with its
associated caption and explanation, and a section group contains a
heading with the elements that follow it up to the
next heading. Page- and document-level containers are
kept for bookkeeping but do not propagate relevance,
since they would connect nearly every element to every
other; singleton groups are likewise inert.

During propagation, reading-order edges are traversed
forward, prose-context edges are traversed from the
paragraph to the visual element, and caption, spatial,
and membership edges are traversed in both directions.
The resulting imbalance in edge counts is handled by
the relation-aware transition matrix in
Section~\ref{sec:supp:propagation}.

\subsection{Cross-Page Relation Proposal}
\label{sec:supp:crr_proposal}

CRR proposes candidate pairs of elements that lie on
different pages of the same document, with at least
one visual element (figure or table) in each pair.
Candidates are generated by five complementary cue
families:
\begin{itemize}
    \item \textbf{Lexical and semantic similarity}
    between element texts, measured with BM25 and
    BGE-M3 embeddings.
    \item \textbf{Visual similarity} between element
    crops, measured with SigLIP~2 embeddings.
    \item \textbf{Textual references}, where a caption
    or passage explicitly refers to a figure or table
    located on another page.
    \item \textbf{Shared numerical anchors}, where two
    elements report the same distinctive values.
    \item \textbf{Table continuations}, where a table
    may continue across a page boundary.
\end{itemize}
Duplicate proposals from different cues are merged,
and each cue family is guaranteed a share of the
shortlist so that frequent similarity-based proposals
do not crowd out rarer but more specific cues. The
shortlist contains at most 1,000 pairs per document
and is rescored by the similarity between the ColPali
patch sets of the two regions; up to 200 pairs are
then submitted for verification.

\subsection{Cross-Page Relation Verification}
\label{sec:supp:crr_verification}

For each shortlisted pair, the VLM verifier
(Qwen3.6-35B-A3B~\cite{qwen36}) receives the two
region crops, their page indices, their element types,
and a bounded amount of their text. It decides whether
the two regions are related and, if so, predicts the
relation type and a confidence. We consider three
relation types: \emph{continuation} (\eg, a table
continuing on a later page), \emph{visual reference}
(\eg, a passage referring to a chart several pages
earlier), and \emph{visual coreference} (\eg, the same
data series shown in two different figures).

Verified relations are retained when their confidence
is at least $0.8$, with at most three links per element
and 25 links per document. For FinRAGBench-V, the
limits are increased to five links per element and 100
per document to accommodate longer financial reports.
Retained links are added to the evidence graph as a
separate relation family. Since CRR is
question-independent, it runs once per document
offline, and no LLM inference is required
at retrieval time.

\section{Graph-Guided Retrieval Details}
\label{sec:supp:graph_retrieval}

\subsection{Multi-View Score Fusion}
\label{sec:supp:fusion}

BM25 and BGE-M3 score elements with text, including
tables through their markdown text. SigLIP~2 scores the
crops of figures, images, and tables. ColPali scores
every element through the patches inside its bounding
box (Equation~\ref{eq:colpali}); since page embeddings
are computed once and shared by all elements on a page,
every element receives a visual score in the context of
its page. Each view is standardized over the elements
it scores in the document (Equation~\ref{eq:fusion}),
and a view that does not score an element contributes
nothing to its fused score. The fused scores are
computed once per question and stored, so the
four-view fusion baseline ranks by exactly the scores
from which graph propagation starts.

\subsection{Relation-Aware Relevance Propagation}
\label{sec:supp:propagation}

Given the fused multi-view relevance scores $s(i)$,
we select the $K$ highest-scoring elements with
positive scores as initial seeds. Their scores
are normalized using softmax to construct the
restart distribution $r_0$. We then extract a
bounded subgraph around these seeds, expanding
relation types in a fixed priority with deterministic
truncation, and limiting the traversal depth and
number of nodes to control retrieval cost.

To propagate relevance across document relationships,
we employ personalized PageRank with a relation-aware
transition matrix. Since spatial proximity edges
are more frequent than other relationships,
a conventional random walk may overemphasize
spatial connections. We therefore normalize
transitions across relation types before
distributing relevance among individual edges:
\begin{equation}
T_{ij} =
\sum_{r\in\mathcal{R}(i)}
\frac{\lambda_r}
{\sum_{t\in\mathcal{R}(i)}\lambda_t}
\frac{c_{ij}^{r}}
{\sum_k c_{ik}^{r}},
\label{eq:supp:transition}
\end{equation}
where $\mathcal{R}(i)$ denotes the outgoing
relation types of node $i$, $\lambda_r$ is
the weight assigned to relation type $r$,
and $c_{ij}^{r}$ is the corresponding edge
confidence. This formulation balances
relevance propagation across different
relation types, allowing less frequent
but informative connections to contribute
without being overwhelmed by spatial edges.

The resulting transition matrix is used in the
personalized PageRank iteration of
Equation~\ref{eq:ppr}, which is run to convergence.
The initial retrieval and graph-propagated rankings
are subsequently combined through reciprocal rank
fusion (Equation~\ref{eq:rrf}). Ranking by propagated
scores alone would favor elements in densely connected
regions of the subgraph, so a strongly matching but
sparsely connected element could fall below weakly
matching elements in a well-connected cluster. Rank
fusion keeps such direct matches near the top while
promoting the elements supported by propagation.

\subsection{Intra-Group Score Refinement}
\label{sec:supp:group_refinement}

After reciprocal rank fusion, we further
refine relevance scores within the visual
and section groups established during graph
construction. This step promotes related
elements that share a common context while
avoiding excessive promotion of weakly
relevant content.

\paragraph{Visual Group Refinement.}
Each visual group contains a figure, table,
or equation together with its associated
caption and explanation. Since these elements jointly
represent a coherent piece of evidence,
we assign every member at least the
highest relevance score within its group:
\begin{equation}
s'(i)=
\max\left(
s_{\mathrm{RRF}}(i),
\max_{j\in G}s_{\mathrm{RRF}}(j)
\right),
\quad i\in G,
\label{eq:supp:visual_refinement}
\end{equation}
where $G$ denotes a visual group and
$s_{\mathrm{RRF}}(i)$ is the relevance
score after reciprocal rank fusion.
This allows a relevant caption to
promote its associated visual element,
and vice versa.

\paragraph{Section Group Refinement.}
Unlike visual groups, section groups
may contain many elements, only some
of which are relevant to the question.
Promoting all members equally could
therefore introduce irrelevant content.
Instead, we select the section heading,
the highest-scoring element and its
two preceding and following neighbors
in reading order, and the $m$
highest-scoring elements within the
section.

Let $\mathcal{T}_m(G)$ denote the
$m$ highest-scoring elements in
section group $G$. We compute a
group-level promotion score as
\begin{equation}
\tau_G =
\frac{1}{|\mathcal{T}_m(G)|}
\sum_{j\in\mathcal{T}_m(G)}
s_{\mathrm{RRF}}(j),
\label{eq:supp:section_score}
\end{equation}
and update the relevance score of
each selected element $i$ according to
\begin{equation}
s'(i)=
\max\left(
s_{\mathrm{RRF}}(i),\tau_G
\right).
\label{eq:supp:section_refinement}
\end{equation}

This strategy retains relevant section
context without promoting every member
indiscriminately: promoted elements are
raised to the mean of the top-$m$ scores,
so section context does not rise above a
strong direct match, and elements that
already score higher keep their own scores.
To prevent cascading score increases across
overlapping groups, all promotion scores are
computed from the original RRF scores before
any updates are applied.

After intra-group score refinement,
auxiliary graph nodes and bridge nodes that are
not eligible retrieval candidates are
excluded from the final layout-level
ranking.

\subsection{Page-Level Retrieval}
\label{sec:supp:page}

The global term $S_{\mathrm{global}}(p)$ in
Equation~\ref{eq:page} is restricted to the ten pages
with the highest whole-page ColPali scores. The local
term $S_{\mathrm{local}}(p)$ captures region-level
evidence and may under-rank a page whose evidence is a
single small table, while the whole-page score has the
opposite bias; fusing the two balances these effects.
The page ranking is computed from the finished
layout-level ranking and never feeds back into it, so
page-level and layout-level results can be read
independently.

\subsection{Retrieval Configuration}
\label{sec:supp:retrieval_config}

We combine the four retrieval signals with
weights of 1 for BM25, BGE-M3, and SigLIP 2,
and 2 for ColPali. Graph propagation uses
50 initial seeds, a maximum of three hops,
and a restart probability of $1 -\alpha=0.15$.
We set the reciprocal rank fusion constant
to $\kappa=60$ and the page-level fusion
weight to $\eta=0.5$. All retrieval
hyperparameters are selected on MMDocIR
and kept fixed across benchmarks.

Table~\ref{tab:supp:hyperparams} summarizes all
retrieval hyperparameters. The ColPali fusion weight is selected
through the sensitivity analysis in
Section~\ref{sec:supp:colpali_weight}. Retrieval is a
single pass without learned parameters, and the bounded
subgraph keeps the per-question cost independent of
document length. Retrieval operations following model
inference run on CPU.

\begin{table}[t]
\centering
\footnotesize
\setlength{\tabcolsep}{4pt}
\caption{Retrieval hyperparameters of \model. All
values are selected on MMDocIR and kept fixed across
benchmarks.}
\label{tab:supp:hyperparams}
\vspace{3pt}
\fitbox{\linewidth}{\begin{tabular}{l l}
\toprule
\textbf{Component} & \textbf{Setting} \\
\midrule
\multicolumn{2}{@{}l@{}}{\textit{Multi-view fusion (Eq.~\ref{eq:fusion})}} \\
View weights $w_m$ & BM25 1, BGE-M3 1, SigLIP~2 1, ColPali 2 \\
\multicolumn{2}{@{}l@{}}{\textit{Propagation (Eqs.~\ref{eq:ppr}, \ref{eq:supp:transition})}} \\
Seeds $K$ & 50 highest-scoring positive elements \\
Subgraph & $\leq$ 3 hops, $\leq$ 100 nodes \\
Restart parameter $\alpha$ & 0.85 \\
Relation weights $\lambda_r$ & 1 (standard), 2 (verified cross-page) \\
\multicolumn{2}{@{}l@{}}{\textit{Rank fusion (Eq.~\ref{eq:rrf})}} \\
Smoothing constant $\kappa$ & 60 \\
\multicolumn{2}{@{}l@{}}{\textit{Intra-group refinement (Eqs.~\ref{eq:supp:visual_refinement}--\ref{eq:supp:section_refinement})}} \\
Visual groups & maximum score within the group \\
Section groups & heading, top element $\pm$ neighbors, top $m$ \\
Top members $m$ & 10 \\
\multicolumn{2}{@{}l@{}}{\textit{Ranking outputs}} \\
Layout ranking length & $\leq$ 200 elements \\
Page fusion weight $\eta$ (Eq.~\ref{eq:page}) & 0.5 \\
Global page prior & top 10 ColPali pages \\
\bottomrule
\end{tabular}}
\end{table}

\section{Answering Details}
\label{sec:supp:answering}

\paragraph{Evidence Package and Answering.}
For downstream QA, we construct an evidence
package containing the top five pages ranked
by page-level relevance and the top ten layout
elements ranked by layout-level relevance.
Following the DocScope~\cite{docscope}
answering setup, we provide the VLM with
the question, low-resolution thumbnails
of all document pages, and our retrieved
evidence package. The thumbnails preserve
global document context, while the evidence
package provides higher-resolution pages
and localized regions to support grounded
answering. We retain the original answering
protocol and augment only its input context.

\paragraph{Input Rendering.}
Following the DocScope~\cite{docscope}
implementation, document thumbnails
are rendered at 72 DPI and retrieved
evidence pages at 144 DPI. The evidence
package is inserted after the document
thumbnails and immediately before the question,
together with a short instruction marking it as
retrieved evidence, so that the model weighs the
package against the full document rather than
following it blindly.
For FinRAGBench-V, we use a fixed
per-page pixel budget. Answer generation
is limited to 16,384 tokens, and failed
or over-length responses are counted
as incorrect.

\paragraph{Answering Protocol.}
DocScope's citation-and-answer protocol~\cite{docscope}
requires factual statements to cite their supporting
page and bounding box. The Qwen3.8-27B baseline uses
the same protocol without an evidence package, and the
four-view fusion baseline builds its package in the
same way from the fused scores without graph
propagation or intra-group refinement. Any difference
between the three settings therefore stems from the
retrieved evidence.

\section{Evaluation Details}
\label{sec:supp:evaluation}

\paragraph{Retrieval Metrics.}
P@$k$ is the fraction of ground-truth pages retrieved
among the top-$k$ pages, and L@$k$ is the fraction of
ground-truth region area covered by the top-$k$
retrieved elements on the gold pages. On MMDocIR,
retrieval results are averaged over its ten domains,
following the official protocol. The same metrics are
applied to DocScope and FinRAGBench-V using their
available evidence annotations.

\paragraph{FinRAGBench-V.}
FinRAGBench-V is evaluated in a single-document
setting, where the target document is provided.
Page-level retrieval and QA use the 539 questions
with gold page annotations, while layout-level
retrieval uses the 91 questions with human-annotated
regions. Therefore, our results measure
within-document retrieval and answering rather
than corpus-level document discovery, and are
not directly comparable with the benchmark's
corpus-level leaderboard.

\paragraph{Answer Evaluation.}
DocScope reports answer accuracy separately for its
679 answerable and 51 unanswerable questions, as well
as overall. For both benchmarks, we follow the official
LLM-as-a-judge procedure, using Qwen3.6-35B-A3B as the
judge. Model configurations, prompts, rendering
settings, and artifact hashes are recorded for
reproducibility.

\paragraph{Question Subsets.}
For the analyses in Sections~\ref{sec:exp:modality}
and~\ref{sec:exp:multipage}, questions are grouped by
the modality of their ground-truth evidence (text,
table, figure, or cross-modal when the evidence spans
several modalities) and by whether the evidence lies
on a single page or on multiple pages. FinRAGBench-V
contains no cross-modal subset.

\section{Additional Experimental Results}
\label{sec:supp:results}

This section reports results that complement the main
paper; results already shown there are not repeated.

\subsection{Additional Ablation Studies}
\label{sec:supp:ablation}

\paragraph{Relation Types.}
Table~\ref{tab:supp:relation_types} keeps a single
relation type in the evidence graph on MMDocIR.
Reading-order edges give the largest individual gain,
improving L@10 from 67.1\% (four-view fusion, no graph)
to 68.7\%, followed by caption (67.7\%) and
prose-context (67.6\%) edges. Section groups alone
leave L@10 unchanged but improve L@5 from 57.2\% to
57.9\%. No single relation type approaches the full
graph (70.7\% L@10), indicating that the relation
types provide complementary connections.

\begin{table}[t]
\centering
\footnotesize
\setlength{\tabcolsep}{5pt}
\caption{Layout recall on MMDocIR when the evidence
graph keeps a single relation type.}
\label{tab:supp:relation_types}
\vspace{3pt}
\fitbox{\linewidth}{\begin{tabular}{l ccc}
\toprule
Relation types & L@1 & L@5 & L@10 \\
\midrule
None (four-view fusion) & 32.8 & 57.2 & 67.1 \\
Reading-order edges only & 32.5 & 58.7 & 68.7 \\
Caption edges only & 32.8 & 57.8 & 67.7 \\
Section groups only & 32.8 & 57.9 & 67.1 \\
Prose-context edges only & \textbf{33.0} & 57.4 & 67.6 \\
\midrule
All (\model) & 32.1 & \textbf{61.1} & \textbf{70.7} \\
\bottomrule
\end{tabular}}
\end{table}

\subsection{Sensitivity to the ColPali Fusion Weight}
\label{sec:supp:colpali_weight}

Table~\ref{tab:supp:colpali_weight} varies the ColPali
weight $w$ in Equation~\ref{eq:fusion} while keeping
the other view weights at 1. Layout recall increases
with $w$ for both four-view fusion and \model, and
\model improves over four-view fusion at every weight,
by 2.9--3.9\% at L@5 and 3.6--4.8\% at L@10. At L@1,
four-view fusion remains slightly higher by at most
0.7\%. The improvement from the evidence graph is thus
not tied to a particular fusion weight.

\begin{table}[t]
\centering
\footnotesize
\setlength{\tabcolsep}{5pt}
\caption{Layout recall on MMDocIR at different ColPali
fusion weights $w$. $w=0$ fuses only three views; the
main paper uses $w=2$.}
\label{tab:supp:colpali_weight}
\vspace{3pt}
\fitbox{\linewidth}{\begin{tabular}{l ccc ccc}
\toprule
& \multicolumn{3}{c}{Four-view fusion} & \multicolumn{3}{c}{\model} \\
\cmidrule(lr){2-4} \cmidrule(lr){5-7}
$w$ & L@1 & L@5 & L@10 & L@1 & L@5 & L@10 \\
\midrule
0 & 28.6 & 48.1 & 55.7 & 28.1 & 51.0 & 60.0 \\
0.5 & 30.1 & 51.5 & 59.9 & 29.9 & 55.0 & 64.7 \\
1 & 31.4 & 54.4 & 63.4 & 30.8 & 58.2 & 67.1 \\
\textbf{2 (ours)} & \textbf{32.8} & \textbf{57.2} & \textbf{67.1} & \textbf{32.1} & \textbf{61.1} & \textbf{70.7} \\
\bottomrule
\end{tabular}}
\end{table}

\subsection{Analysis by Evidence Modality}
\label{sec:supp:modality}

Table~\ref{tab:supp:modality} extends
Table~\ref{tab:modality} to DocScope and FinRAGBench-V.
\model achieves the best layout recall for every
modality on both benchmarks. As on MMDocIR, the largest
gains over four-view fusion occur on figures, where
L@10 improves by 6.6\% on DocScope and 16.1\% on
FinRAGBench-V. Text-only retrievers nearly fail to
localize figures on both benchmarks (at most 5.5\%
L@10), since figures carry no parsed text, which
highlights the importance of the visual views and of
graph connections to captions and surrounding text.

\begin{table}[t]
\centering
\footnotesize
\setlength{\tabcolsep}{4pt}
\caption{Layout-level recall (L@10) by evidence modality
on DocScope and FinRAGBench-V. FinRAGBench-V has no
cross-modal subset.}
\label{tab:supp:modality}
\vspace{3pt}
\fitbox{\linewidth}{\begin{tabular}{l cccc ccc}
\toprule
& \multicolumn{4}{c}{DocScope} & \multicolumn{3}{c}{FinRAGBench-V} \\
\cmidrule(lr){2-5} \cmidrule(lr){6-8}
Method & Text & Table & Figure & Cross & Text & Table & Figure \\
\midrule
BM25~\cite{bm25} & 37.9 & 30.6 & 3.0 & 20.5 & 28.3 & 25.8 & 5.3 \\
BGE-M3~\cite{bge} & 44.0 & 49.6 & 2.0 & 27.3 & 25.7 & 23.5 & 5.5 \\
Four-view fusion & 51.0 & 63.5 & 44.8 & 49.7 & 36.2 & 53.6 & 42.8 \\
\textbf{\model} & \textbf{51.5} & \textbf{66.0} & \textbf{51.4} & \textbf{53.1} & \textbf{38.3} & \textbf{54.1} & \textbf{58.9} \\
\bottomrule
\end{tabular}}
\end{table}

\subsection{Analysis by Evidence Distribution}
\label{sec:supp:multipage}

Table~\ref{tab:supp:pages} extends Table~\ref{tab:pages}
to DocScope and FinRAGBench-V. \model achieves the best
layout recall on both single-page and multi-page
questions of both benchmarks. Compared with the four-view fusion setting,
L@10 improves by 4.9\% and 1.3\% on single-page and
multi-page DocScope questions, and by 6.9\% and 8.6\%
on single-page and multi-page FinRAGBench-V questions,
correspondingly.

\begin{table}[t]
\centering
\footnotesize
\setlength{\tabcolsep}{5pt}
\caption{Layout-level recall (L@10) on single-page and
multi-page questions of DocScope and FinRAGBench-V.}
\label{tab:supp:pages}
\vspace{3pt}
\fitbox{\linewidth}{\begin{tabular}{l cc cc}
\toprule
& \multicolumn{2}{c}{DocScope} & \multicolumn{2}{c}{FinRAGBench-V} \\
\cmidrule(lr){2-3} \cmidrule(lr){4-5}
Method & Single & Multi & Single & Multi \\
\midrule
BM25~\cite{bm25} & 33.1 & 23.0 & 19.1 & 12.2 \\
BGE-M3~\cite{bge} & 44.2 & 32.2 & 17.7 & 10.7 \\
Four-view fusion & 66.1 & 47.1 & 46.6 & 31.2 \\
\textbf{\model} & \textbf{71.0} & \textbf{48.4} & \textbf{53.5} & \textbf{39.8} \\
\bottomrule
\end{tabular}}
\end{table}

\end{document}